\documentclass[12pt]{article}
\begin{document}
\begin{center}
\large\bf{Derivative Expansion of the Effective Action for Massless\\
Scalar Electrodynamics in Arbitrary Gauge}
\end{center}
\vspace{2cm}
\begin{center}
D.G.C. McKeon\\
Department of Applied Mathematics\\
University of Western Ontario\\
London\\
CANADA\\
N6A 5B7
\end{center}
\vspace{1cm}
email:  dgmckeo2@uwo.ca\\
Tel: (519)661-2111, ext. 88789\\
Fax: (519)661-3523\vspace{3cm}\\

\section{Abstract}
 It is shown how operator regularization can be used to obtain an expansion of the effective action in powers of derivatives of the background field. This is applied to massless scalar electrodynamics to find the one-loop corrections to the kinetic terms associated with both the scalar and vector fields in arbitrary gauge. This allows us to examine the radiatively induced masses arising in this model.
 
\section{Introduction}

The expansion of the effective action in powers of derivatives of the background field 
has in the past received considerable attention, both when the background field 
is a scalar [1-9] and vector [10-16] field. Examination of the effective action when the background field is a constant scalar field has led to the idea of radiatively induced spontaneous symmetry breaking [1-4].

In this paper we first demonstrate how this derivative expansion can be derived when the effective action is computed using operator regularization [17-20]. This regulating technique provides a way of avoiding divergences when evaluating  radiative corrections without having to introduce any modifications to the initial classical action. The approach used is similar to that employed in ref. [21].

Using this technique, we then compute the one loop radiative corrections to the kinetic terms in the effective action for massless scalar electrodynamics. This involves expanding up to the second derivative of the background field when examining the two point functions for the scalar and vector fields. This is done in a class of gauges similar to the $R_\xi$ gauges [22-23].

\section{The Derivative Expansion}

To illustrate our approach, let us consider a simple scalar model with action
$$S = \int d^4x \left[-\frac{1}{2} (\partial\phi)^2 - \frac{1}{2} m^2\phi^2 - \frac{1}{6}\mu\phi^3 - \frac{1}{24} \lambda\phi^4\right].\eqno(1)$$
The field $\phi$ is then split into the sum of a background field $f$ and a quantum field $h$ so that
$$\phi = f + h.\eqno(2)$$
If now $\partial = ip$, the contribution to $S$ that is bilinear in $h$ is given by
$$S_2 = \int d^4x \left[ - \frac{1}{2} h\left( p^2 + m^2 + \mu f + \frac{1}{2} \lambda f^2\right)h\right]\eqno(3)$$
so that at one loop order the effective action is
$$iS^{(1)} = \ln det^{-1/2} \left( p^2 + m^2 + \mu f + \frac{1}{2} \lambda f^2\right).\eqno(4)$$
In operator regularization [17-20], the zeta function [24-26] is used to regulate functional logarithms,
$$\ln H = -\frac{d}{ds}\left|\raisebox{-2ex}{\scriptsize{0}} H^{-s}\right. .\eqno(5)$$
This allows us to regulate $iS^{(1)}$ in eq. (4),
$$iS^{(1)} = \frac{1}{2} \frac{d}{ds}\left|\raisebox{-2ex}{\scriptsize{0}} \int_0^\infty d it (it)^{s-1} tr \exp(-i Ht)\right.\eqno(6)$$
where $H$ is the operator appearing in eq. (4). This entails using $tr \ln H = \ln det H$ and 
$$\Gamma (s)H^{-s} = \int_0^\infty d it (it)^{s-1} \exp (-i Ht).\eqno(7)$$

An expansion due to Schwinger [10]
$$tr\, e^{-i(H_0 + H_1)t} = tr \left(e^{-iH_0t} + (-it) e^{-iH_0t} H_1
+ \frac{(-it)^2}{2} \int_0^1 du\, e^{-i(1-u)H_0t} H_1 e^{-iuH_0t}H_1 + \ldots \right)\eqno(8)$$
can now be used to expand the exponential appearing in eq. (6) in powers of $f$.

In making a derivative expansion of $iS^{(1)}$, we first make the replacement
$$f \rightarrow v + f\eqno(9)$$
where $v$ is a constant; this permits one to retain all contributions that are independent of derivatives in the background field. With this, we find that
$$H_0 = p^2 + m^2 + \mu v + \frac{\lambda v^2}{2} \eqno(10)$$
$$H_1 = (\mu + \lambda v) f +  \frac{\lambda f^2}{2}. \eqno(11)$$

Keeping terms in eq. (8) that are second order in $f$, we find that
$$iS^{(1)}_2 = \frac{1}{2} \frac{d}{ds}\left|\raisebox{-2ex}{\scriptsize{0}} \frac{\kappa^{2s}}{\Gamma(s)} \int_0^\infty d\, it (it)^{s-1} tr\left\lbrace (-it)e^{-i(p^2 +m^2+\mu v+\frac{\lambda v^2}{2})t}\left((\mu + \lambda v)f + \frac{\lambda f^2}{2}\right)\right.\right.\nonumber$$
$$ \left. + \frac{(-it)^2}{2} \int_0^1 du\, e^{-i(1-u)(p^2 + m^2 + \mu v + \frac{\lambda v^2}{2})t} (\mu + \lambda v)f
e^{-iu\left(p^2 + m^2 + \mu v + \frac{\lambda v^2}{2}\right) t} (\mu + \lambda v) f\right\rbrace .\eqno(12)$$
The factor of $\kappa^{2s}$ has been inserted to ensure that $t$ is dimensionless.

The first functional trace we compute in eq. (12) is
$$T_1 = tr \,e^{-ip^2t} f^2 = \int dx \int dp <x| e^{-ip^2t} |p><p| f^2|x>.\eqno(13)$$
If now $|x>$ and $|p>$ are the usual position and momentum eigenstates so that 
$$(2\pi)^2 <x|p> = e^{ip\cdot x}\eqno(14)$$
then eq. (13) becomes
$$T_1 = \int dx \int \frac{dp}{(2\pi)^4}\, e^{-ip^2t}f^2(x)
= \frac{i}{(4\pi it)^2} \int dx f^2(x) .\eqno(15)$$
For the other functional trace in eq. (12),
$$T_2 = tr\, e^{-i(1-u)p^2t} f e^{-iup^2t}f\nonumber$$
$$= \int dx \int dy\, \delta (x-y) \int dp\, dq \int dz\,
<x|e^{-i(1-u)p^2t}|p><p|f|z><z|e^{-iup^2t}|q><q|f|y>\nonumber$$
$$= \int dx\, dy \delta (x-y) \int dp \,dq \int dz\, \frac{e^{ip\cdot(x-z)}e^{iq\cdot(z-y)}}{(2\pi)^8}\eqno(16)$$
$$ e^{-i[(1-u)p^2+uq^2]t} f(z) f(y).\nonumber$$
We now expand $f(z)$ about $y$, stopping at the second derivative, so that
$$\int dq \,e^{-iuq^2t} f(z) e^{iq\cdot(z-y)}\eqno(17)$$
$$=\int dq\, e^{-iuq^2t} \left[f(y) + (z-y)^\lambda f_{,\lambda}(y)
+ \frac{1}{2} (z-y)^\lambda (z-y)^\sigma f_{,\lambda\sigma}(y)\right] e^{iq\cdot (z-y)}.\nonumber$$
Noting that
$$(z-y)^\lambda e^{iq\cdot (z-y)} = -i\,\frac{\partial}{\partial q^\lambda} e^{iq\cdot (z-y)}\eqno(18)$$
and integrating by parts with respect to $q$, eq. (17) becomes
$$= \int dq\, e^{iq\cdot (z-y)}\left( f(y) + \left(i \frac{\partial}{\partial q}\right)_\lambda f_{,\lambda}(y)\right. \nonumber$$
$$\left. +\frac{1}{2} \left(i \frac{\partial}{\partial q}\right)_\lambda \left(i \frac{\partial}{\partial q}\right)_\sigma f_{,\lambda \sigma}(y)\right)e^{-iuq^2t}
\nonumber$$
$$= \int dq\, e^{iq\cdot (z-y)}\left[ f(y) + (2uq^\lambda t) f_{,\lambda}(y)\right. \nonumber$$
$$\left. +\frac{1}{2}\left((2uq^\lambda t)(2uq^\sigma t) + 2iut g^{\lambda\sigma}\right) f_{,\lambda\sigma}(y)\right] e^{-iuq^2t}.\eqno(19)$$

Substitution of eq. (19) into eq. (16) leads to some standard integrals; we obtain after integration over $z$, $y$, $q$ then $p$ in succession,
$$T_2 = \int dx \frac{i}{(4\pi it)^2} \left(f^2(x) + (it)u(1-u)f(x)\partial^2 f(x)\right).\eqno(20)$$
Together, eqs. (15) and (20) bring eq. (12) into the form
$$iS^{(1)}_2 = \frac{1}{2} \frac{d}{ds}\left|\raisebox{-2ex}{\scriptsize{0}}\; \frac{\kappa^{2s}}{\Gamma(s)} \int_0^\infty d it (it)^{s-1} \frac{i}{(4\pi it)^2} e^{-i(m^2 +\mu v +\frac{1}{2} \lambda v^2)t}\right. \eqno(21)$$
$$\left[\int dx (-it)\left((\mu + \lambda v) f(x) + \frac{1}{2} \lambda f^2 (x)\right)\right.\nonumber$$
$$\left. + \frac{(-it)^2}{2} \int_0^1 du (\mu + \lambda v)^2 \left(f^2 (x) + (it) u(1-u)f(x)\partial^2 f(x)\right)\right],\nonumber$$
which by eq. (7) becomes
$$= \frac{i}{32\pi^2} \frac{d}{ds}\left|\raisebox{-2ex}{\scriptsize{0}}\; \frac{\kappa^{2s}}{\Gamma(s)} \int dx\right.\nonumber$$ 
$$\left[-\left( (\mu + \lambda v) f(x) + \frac{1}{2} \lambda f^2(x)\right)\Gamma(s-1)
\left(m^2 + \mu v + \frac{1}{2} \lambda v^2 \right)^{1-s}\right.\nonumber$$ 
$$+ \frac{1}{2} (\mu + \lambda v)^2 f^2 (x)\Gamma (s)
\left(m^2 + \mu v + \frac{1}{2} \lambda v^2 \right)^{-s}\nonumber$$ 
$$\left. + \frac{1}{12} (\mu + \lambda v)^2 f (x)\partial^2 f(x)
\Gamma (s+1)\left(m^2 + \mu v + \frac{1}{2} \lambda v^2\right)^{-1-s}\right].\eqno(22)$$
Since $\Gamma(x + 1) = x\Gamma (x)$, eq. (22) becomes
$$iS_2^{(1)} = \frac{i}{32\pi^2} \int dx \left\lbrace \left[ (\mu + \lambda \sigma )f(x) + \frac{1}{2} \lambda f^2 (x)\right]\left[m^2 + \mu v + \frac{1}{2} \lambda\sigma^2\right]
\left[ 1 - \ln \left( \frac{m^2 + \mu v + \frac{1}{2} \lambda v^2}{\kappa^2}\right)\right]\right.\nonumber$$ 
$$\left. - \frac{1}{2} (\mu + \lambda v)^2 f^2(x)
\ln \left[ \frac{m^2 + \mu v + \frac{1}{2} \lambda v^2}{\kappa^2}\right] + \frac{1}{12} \frac{(\mu + \lambda v)^2 f(x)\partial^2 f(x)}{(m^2 + \mu v + \frac{1}{2} \lambda v^2)}\right\rbrace\, .\eqno(23)$$
This agrees with the result of ref. [5-9].

The terms in the classical actions that are linear and bilinear in $f$ are
$$S_2^{(0)} = \int dx\left[-\left(m^2v + \frac{1}{2} \mu v^2 + \frac{\lambda v^3}{6}\right) f -\frac{1}{2} f\left(p^2 + m^2 + \mu v + \frac{1}{2} \lambda v^2\right)f\right].\eqno(24)$$
We can now write $\left(S_2^{(0)} + S_2^{(1)}\right)$ in the form
$$\left(S_2^{(0)}+ S_2^{(1)}\right) = \int dx\left[
Z(v) f(x) \partial^2 f(x) - V^\prime(v) f(x) - V^{\prime\prime}(v) f^2 (x)\right]
\eqno(25)$$
so that the vacuum expectations value of $\phi$ is given by $v$ where [2]
$$V^\prime (v) = 0\eqno(26)$$
and the radiatively corrected mass for $\phi(x)$ is
$$M^2 = V^{\prime\prime} (v)/Z(v) .\eqno(27)$$
We now apply this technique to massless scalar electrodynamics.

\section{Massless Scalar Electrodynamics}

The classical action we consider is given by
$$S_{c1} = \int dx \left[ - (\partial + ieV)_\mu \phi^* (\partial - ieV)^\mu \phi - \lambda \left(\phi^* \phi\right)^2 - \frac{1}{4} \left(\partial_\mu V_\nu - \partial_\nu V_\mu\right)^2\right].\eqno(28)$$
If now the scalar field $\sqrt{2}\phi$ is written as the
sum of a background field $f$ and quantum field $h$,
$$\sqrt{2}\phi = f + h\eqno(29)$$
the Lagrangian possesses the infinitesimal gauge invariance
$$\delta V_\mu = 0\;\;\;\;\,\,\,\,\delta f = 0 \;\;\;\;\,\,\,\, \delta h = ie(f + h)\Omega .\eqno(30)$$
This gauge invariance is broken by a modified form of the $R_\xi$ gauge [22,23] so  we supplement $S_{c1}$ with
$$S_{gf} = \int dx \left( - \frac{1}{2\alpha}\right) \left(\partial \cdot V + \frac{ie\alpha}{2} (f^*h-fh^*)\right)^2,\eqno(31)$$
which necessitates a ghost action [22,23]
$$S_{gh} = \int dx \,\overline{c} \left[\partial^2 - \frac{1}{2} e^2\alpha \left(2f^* f + f^* h + fh^*\right)\right]c.\eqno(32)$$
The action $S^{(0)} = S_{c1} + S_{gf} + S_{gh}$ has a portion bilinear in the quantum fields $V_\mu$, $h$, $c$ and $\overline{c}$ given by
$$S^{(0)}_2 = -\overline{c} H_g c - \frac{1}{2} \Phi^T H_B \Phi\eqno(33)$$
where
$$H_g = p^2 + e^2 \alpha\left(f_1^2 + f_2^2\right)\eqno(34)$$
$$H_B = \left(
\begin{array}{ccc}
p^2 + 3\lambda f_1^2 + (\lambda + \alpha e^2)f_2^2 &
(2\lambda - \alpha e^2)f_1f_2 &
-2e(\partial_\nu f_2)\\
(2\lambda - \alpha e^2)f_1f_2 &
p^2 + (\lambda + \alpha e^2)f_1 + 3\lambda f_2 &
2e(\partial_\nu f_1)\\
-2e(\partial_\mu f_2) & 2e(\partial_\mu f_1) & \!\!p^2(T + \frac{1}{\alpha} L)_{\mu\nu}\\
 & &
\;\;\;+ e^2\left(f_1^2 + f_2^2\right)g_{\mu\nu}
\end{array}\right)\eqno(35)$$
$$\Phi^T = \left(h_1, h_2, V_\mu\right)\eqno(36)$$
and we have set $f = f_1 + if_2$, $h = h_1 + ih_2$.

The replacement $f_1 \rightarrow v + f_1$ as in eq. (9) leads to
$H_g = H_g^0 + H_g^1 + H_g^2$ and $H_B = H_B^{0}+ H_B^1 + H_B^2$ with
$$H_g^0 = p^2 + e^2 \alpha v^2\eqno(37)$$
$$H_g^1 = 2e^2 \alpha v f,\eqno(38)$$
$$H_g^2 = e^2 \alpha \left(f_1^2 + f_2^2\right)\eqno(39)$$
and
$$H_B^0 = \left(
\begin{array}{ccc}
p^2 + 3\lambda v^2 & 0 & 0 \\
0 & p^2 + (\lambda + \alpha e^2)v^2 & 0 \\
0 & 0 & p^2(T + \frac{1}{\alpha} L)_{\mu\nu} + e^2v^2g_{\mu\nu}\end{array}
\right)\eqno(40)$$
$$H_B^1 = \left(
\begin{array}{ccc}
6\lambda v f_1 & 2\left(\lambda - \frac{1}{2} \alpha e^2\right)vf_2 & -2e(\partial_\nu f_2)\\
2\left(\lambda - \frac{1}{2} \alpha e^2\right)vf_2 & 
2\left(\lambda + \alpha e^2\right)vf_1 & 2e\left(\partial_\nu f_1\right)\\
-2e\left(\partial_\mu f_2\right) & 2e\left(\partial_\mu f_1\right) &
2e^2vf_1g_{\mu\nu}
\end{array}\right)\eqno(41)$$
$$H_B^2 = \left(\begin{array}{ccc}
3\lambda f_1^2 + \left(\lambda +\alpha e^2\right)f_2^2 & (2\lambda - \alpha e^2)f_1f_2 & 0\\
\left(2\lambda - \alpha e^2\right)f_1f_2 & \left(\lambda + \alpha e^2\right)f_1^2 + 3\lambda f_2^2 & 0\\
0 & 0 & e^2\left(f_1^2 + f_2^2\right)g_{\mu\nu}
\end{array}\right).\eqno(42)$$
(Derivatives appearing in eq. (41) act only on the classical field; i.e. $\left(\partial_\mu f\right) \equiv \left(\frac{\partial}{\partial x^\mu} f(x)\right)$.)  We have used a complete set of orthogonal projection operators $T_{\mu\nu} = g_{\mu\nu} - \partial_\mu\partial_\nu/\partial^2$, $L_{\mu\nu} = \partial_\mu\partial_\nu /\partial^2$.

The one loop contribution to the effective action is now given by $S^{(1)} = S^{(1)}_g + S^{(1)}_B$ where
$$iS_g^{(1)} = \ln det H_g\, ,\,\,\,\, iS_B^{(1)} = \ln det^{-1/2} H_B
\, . \eqno(43)$$
Through using eqs. (5,7,8), we find that to second order in $f_1$ and $f_2$,
$$iS^{(1)} = \frac{d}{ds}\left|\raisebox{-2ex}{\scriptsize{0}}\; \frac{\kappa^{2s}}{\Gamma(s)} \int_0^\infty dit(it)^{s-1} tr\left\lbrace -\left[ (-it)e^{-iH_g^0t}\left(H_g^1 + H_g^2\right)
+ \frac{(-it)^2}{2} \int_0^1 du\, e^{-i(1-u)H_g^0t} H_g^1 e^{-iuH_g^0t} H_g^1\right]\right. \right. \nonumber$$
$$\left. + \frac{1}{2} \left[ (-it) e^{-iH_B^0t}\left(H_B^1 + H_B^2\right)+ \frac{(-it)^2}{2} \int_0^1 du\, e^{-i(1-u)H_B^0 t}
H_B^1 e^{-iuH_B^0t}H_B^1\right]\right\rbrace .\eqno(44)$$
Upon substitution of eqs. (37-42) into eq. (44), we are left with functional traces of the form of $T_1$ and $T_2$ in eqs. (15) and (20) respectively, as well as
$$T_3 = tr\left( e^{-i(1-u)p^2t}\, \partial_\mu f\left(e^{-iup^2(T+\frac{1}{\alpha}L)t}\right)^{\mu\nu} \partial_\nu f\right)\eqno(45)$$
and
$$T_4 = tr\left(\left( e^{-i(1-u)p^2\left(T + \frac{1}{\alpha} L\right)t} \right)_{\mu\nu} f\left(e^{-iup^2(T+\frac{1}{\alpha}L)t}\right)^{\nu\mu}  f\right).\eqno(46)$$
Since $T$ and $L$ are a complete set of orthogonal projection operators, it follows that
$$\left(e^{-ip^2\left(T + \frac{1}{\alpha} L\right)t}\right)_{\mu\nu} = e^{-ip^2t} T_{\mu\nu} + e^{-ip^2t/\alpha}L_{\mu\nu}.\eqno(47)$$
Through insertion of complete sets of states as was done in eq. (16), we find then that
$$T_3 = \int dx\, dy \delta(x-y) \int dp\, dq \int dz \frac{e^{ip\cdot (x-z)}e^{iq\cdot (z-y)}}{(2\pi)^8}\nonumber$$
$$\left[e^{-i(1-u)p^2t}\left(g_{\mu\nu} - \frac{p_\mu p_\nu}{p^2}\right) + e^{-i(1-u)p^2t/\alpha}\frac{p_\mu p_\nu}{p^2}
\right]
\partial^\mu f(z) e^{-iuq^2t}\partial^\nu f(y).\eqno(48)$$
Since we are only interested in retaining contributions to $S^{(1)}$ that have at most two derivatives of the background field, in eq. (48) we can make the approximation $\partial^\mu f(z) \cong \partial^\mu f(y)$. The integrals over $z$, $y$ then $q$ lead to
$$T_3 = \int dx \int \frac{dp}{(2\pi)^4} \left(\partial_\mu f (x) \partial^{\mu} f(x)\right) \left(e^{-ip^2t}\left(g_{\mu\nu} - \frac{p_\mu p_\nu}{p^2}\right)
+e^{-i\left((1-u)+u/\alpha\right)p^2t}\left(\frac{p_\mu p_\nu}{p^2}\right)
\right)\eqno(49)$$
$$= \int dx \frac{(\partial f)^2}{4}
\left(\frac{i}{(4\pi it)^2}\right) \left(3 + \frac{1}{\left(1-u + \frac{u}{\alpha}\right)^2}\right).\eqno(50)$$

The trace $T_4$ can also be reduced by the insertion of complete sets of states as in eq. (16), leaving us with
$$T_4 = \int dx\, dy \delta(x-y) \int dp \,dq \int dz\,\, e^{ip\cdot (x-z)}e^{iq\cdot(z-y)}\nonumber$$
$$\left[e^{-i(1-u)p^2t}\left(g_{\mu\nu} - \frac{p_\mu p_\nu}{p^2}\right) + e^{-iup^2t}\frac{p_\mu p_\nu}{p^2}\right] f(z)\eqno(51)$$
$$\left[e^{-iuq^2t}\left(g^{\mu\nu} - \frac{q_\mu q_\nu}{q^2}\right) + e^{-iuq^2t/\alpha}\frac{q_\mu q_\nu}{q^2}\right] f(y).\nonumber$$
We now can expand $f(z)$ about $y$ up to the second derivative
$$f(z) = f(y) + (z-y)^\lambda f_{,\lambda}(y) + \frac{1}{2} (z-y)^\lambda (z-y)^\sigma f_{,\lambda\sigma}(y)\eqno(52)$$
and then in analogy with eqs. (17-19) we are able to recast $T_4$ in a form that makes it possible to evaluate all integrals appearing in eq. (51). We are left with
$$T_4 = \frac{i}{(4\pi it)^2} \int dx \left\lbrace f^2 (3 + \alpha^2) + it f \partial^2 f  \left[ (3 + \alpha)u(1-u) + \frac{3}{4} \frac{\alpha^2(1-u) + u - \alpha}{\alpha (1-u) + u}\right]\right\rbrace . \eqno(54)$$

When the functional traces of eqs. (15), (20), (50) and (54) are all taken into account, then from eq. (44)
$$i S_g^{(1)} = \frac{i}{(4\pi)^2} \left\lbrace 2\alpha^2 e^4 v^3 f_1 \left(\ln \left(\frac{\alpha e^2 v^2}{\kappa^2}\right)-1\right)\right. \eqno(55)$$
$$\left. + \alpha^2 e^4 v^2 \left[f_1^2 \left(\ln \left(\frac{\alpha e^2 v^2}{\kappa^2}\right)-1\right) + f_2^2
\left(3\ln \left(\frac{\alpha e^2 v^2}{\kappa^2}\right)-1\right)\right] - \frac{1}{3} \alpha e^2 f_1 \partial^2 f_1\right\rbrace\nonumber$$
and
$$i S_B^{(1)} = \frac{i}{2(4\pi)^2} \int_0^1 du
\left\lbrace -\left[(6\lambda)(3\lambda v^2)
\left(\ln \frac{3\lambda v^2}{\kappa^2}- 1 \right)\right.\right. \eqno(56)$$
$$ + 2(\lambda +  \alpha e^2 )\left( \left(\lambda + \alpha e^2\right)v^2\right)
\left(\ln \frac{\left(\lambda + \alpha e^2\right) v^2}{\kappa^2}-1\right)\nonumber$$
$$ \left. + (3 + \alpha^2) (2e^2) (e^2 v^2)\left(\ln \frac{e^2 v^2}{\kappa^2}-1\right)\right] (v f_1)\nonumber$$
$$-\left[\left( 3\lambda f_1^2 + (\lambda + \alpha e^2)f_2^2\right) (3\lambda v^2) \left(\ln \frac{3\lambda v^2}{\kappa^2} -1 \right)\right.\nonumber$$
$$+ \left(\left(\lambda + \alpha e^2\right)f_1^2 + 3\lambda f_2^2\right)\left(\left(\lambda + \alpha e^2\right) v^2\right)\left(\ln \frac{(\lambda + \alpha e^2)v^2}{\kappa^2} - 1 \right)\nonumber$$
$$\left. + (3 + \alpha^2)e^2 \left(f_1^2 + f_2^2\right)\left(e^2 v^2\right)\left(\ln\frac{e^2v^2}{\kappa^2} - 1 \right)\right]\nonumber$$
$$+\frac{1}{2} \left[(6\lambda v)^2 \left(-f_1^2 \ln \frac{3\lambda v^2}{\kappa^2} + \frac{u(1-u)f_1\partial^2f_1}{3\lambda v^2}\right)\right.\nonumber$$
$$\left(2 (\lambda + \alpha e^2)v\right)^2 \left(-f_1^2 \ln \frac{(\lambda + \alpha e^2)v^2}{\kappa^2} + \frac{u(1-u)f_1\partial^2f_1}{(\lambda + \alpha e^2)v^2}\right)\nonumber$$
$$+ (2e^2v)^2 \left( -(3 + \alpha^2)f_1^2 \ln \frac{e^2v^2}{\kappa^2} +
\left(\frac{3}{4} \frac{\alpha^2(1-u)+u-\alpha}{\alpha(1-u)+u} + (3+\alpha)(1-u)\right)
\frac{f_1\partial^2 f_1}{e^2v^2}\right)\nonumber$$
$$+ 2(2\lambda - \alpha e^2)^2 v^2 \left( 
-f_2^2 \ln \frac{(1-u)(3\lambda v^2)+u(\lambda +\alpha e^2)v^2}
{\kappa^2} +
\frac{u(1-u)f_2\partial^2f_2}{(1-u)(3\lambda v^2)+u(\lambda + \alpha e^2)v^2}
\right)\nonumber$$
$$+ 2e^2 \left( 3 + \frac{1}{(1-u+\frac{u}{\alpha})^2}\right)
\left(-(\partial f_2)^2 \ln\left(\frac{(1-u)(3\lambda v^2)+ue^2v^2}{\kappa^2}\right)\right.\nonumber$$
$$\left.\left.\left. -(\partial f_1)^2 \ln \left( \frac{(1-u)(\lambda + \alpha e^2)v^2 + ue^2v^2}{\kappa^2}\right)\right)\right]\right\rbrace .\nonumber$$
Integrals over the Schwinger parameter $u$ in eq. (56) are straightforward, albeit tedious:
$$\int_0^1 du\, u(1-u) = \frac{1}{6}\eqno(57)$$
$$\int_0^1 du\, \ln(A(1-u)+Bu) = \frac{B\ln B - A\ln A}{B -A} - 1 \eqno(58)$$
$$\int_0^1 du\, \frac{\alpha^2 (1-u) + u - \alpha}{\alpha (1-u) + u} = (1 + \alpha) + \frac{2 \alpha \ln \alpha}{1 - \alpha} \eqno(59)$$
$$\int_0^1 du\, \frac{u (1-u)}{A(1-u) + Bu} = \frac{1}{(B - A)^3}
\left[ \frac{1}{2} (B^2 - A^2) - AB \ln \frac{B}{A}\right] \eqno(60)$$
$$\int_0^1 du\, \frac{\ln(A(1-u) + Bu)}{(1-u+ \frac{u}{\alpha})^2} = \frac{\alpha}{1 - \alpha}
\left[ \ln A - \alpha \ln B + 
\frac{\alpha(A-B)}{A-\alpha B} \ln \left( \frac{\alpha B}{A}\right)\right].\eqno(61)$$
When eqs. (57-61) are used, eqs. (55,56) provide the one loop contribution to $Z$, $V^\prime$ and $V^{\prime\prime}$ of eq. (25) in the gauge of eq. (31).

We now turn to the self energy of the vector field, computing the expansion of the one loop correction to the effective potential that is bilinear in the background vector field and second order in its derivatives. This entails decomposing $V_\mu$ and $\sqrt{2} \phi$ in eq. (28) into a sum of classical ($A_\mu$, $v$) and quantum ($a_\mu$, $h_1 + ih_2$) parts so that
$$V_\mu (x) = A_\mu(x) + a_\mu(x)\eqno(62)$$
$$\sqrt{2} \phi (x) = v + h_1 (x) + ih_2(x).\eqno(63)$$
The classical action now contains the bilinear contributions
$$S_{c1}^{(2)} = -\frac{1}{2} (h_1, h_2, a_\mu)\eqno(64)$$
$$\left(
\begin{array}{ccc}
p^2 + 3\lambda v^2 + e^2A^2 &
-ie(p_\mu A^\mu + A^\mu p_\mu) & 2e^2 vA_\nu \\
ie(p_\mu A^\mu + A^\mu p_\mu) & 
p^2 + (\lambda + \alpha e^2)v^2 + e^2A^2 & 0 \\
2e^2v A_\mu & 0 & p^2(T + \frac{a}{\alpha} L)_{\mu\nu}\\
 & &\;\;\; + g_{\mu\nu}e^2v^2 \end{array}\right)\left(\begin{array}{c}
h_1\\
h_2\\
a_\nu \end{array}\right).\nonumber$$
The only contribution to the one loop classical action now is given by
$$iS^{(1)}_A = \ln det^{-1/2} H_A\eqno(65)$$
where $H_A$ is the operator appearing in eq. (64). Again regulating this expression through use of eq. (6) and expanding to second order in $A_\mu$ by use of eq. (7), we find that
$$iS_A^{(1)} = \frac{1}{2} \frac{d}{ds}\left|\raisebox{-2ex}{\scriptsize{0}}\; \frac{\kappa^{2s}}{\Gamma(s)} \int_0^\infty d\,it (it)^{s-1} tr\left\lbrace (-it)e^{-ip^2t}
\left(e^{-i(3\lambda v^2)t} + e^{-i(\lambda +\alpha e^2)v^2t}\right)(e^2A^2)\right.\right.\eqno(66)$$
$$+\frac{(-it)^2}{2} \int_0^1 du \left[2e^2 e^{-i[(1-u)(\lambda + \alpha e^2) +u (3\lambda)]v^2 t}
\left( e^{-i(1-u)p^2t} (p \cdot A + A \cdot p) e^{-iup^2t} (p \cdot A + A \cdot p)\right)\right.\nonumber$$
$$+2(2e^2 v)^2 e^{-i[(1-u)e^2+u(3\lambda)]v^2t}
\left.\left. \left(e^{-i(1-u)p^2(T + \frac{1}{\alpha} L)t}\right)_{\mu\nu} A^\mu e^{-iup^2t} A^\nu\right]\right\rbrace .\nonumber$$
In addition to a functional trace of the form of $T_1$ appearing in eq. (13), we have in eq. (66) two more functional traces
$$T_5 = tr\, e^{-i(1-u)p^2t} (p \cdot A + A \cdot p) e^{-iup^2t} (p \cdot A + A \cdot p)\eqno(67)$$
and
$$T_6 = tr\left[\left( e^{-i(1-u)p^2t} T_{\mu\nu} + e^{-i(1-u)p^2t/\alpha}L_{\mu\nu}\right) A^\mu e^{-iup^2t} A^\nu\right] .\eqno(68)$$
For $T_5$, we again insert complete sets of eigenstates, as in eq. (16), so that
$$T_5 = \int dx \,dy \delta (x-y)\left[\int dp\, dq \int dz \frac{e^{ip\cdot(x-z)}e^{iq\cdot(z-y)}}{(2\pi)^8}
\left(e^{-i[(1-u)p^2+uq^2]t} (p+q)_\mu q_\nu A^\mu (z) A^\nu (y) \right)\right.\eqno(69)$$
$$\left. + \int dp\,dq\,dr \int dz\,dw \frac{e^{ip\cdot (x-z)iq\cdot (z-w)}e^{ir\cdot (w-y)}}{(2\pi)^{12}} \left(e^{-i[(1-u)p^2+uq^2]t} (p+q)_\mu r_\nu A^\mu (z) A^\nu (w) \right)\right].\nonumber$$
In eq. (69), we effect the expansions
$$A^\mu (z) A^\nu (y) = \left[A^\mu (y) + (z - y)^\lambda A^\mu_{,\lambda} (y) + \frac{1}{2} (z-y)^\lambda (z-y)^\sigma A^{\mu}_{,\lambda\sigma} (y)\right]A^\nu(y)\eqno(70)$$
and
$$A^\mu (z) A^\nu (w) = \left[A^\mu (x) + (z - x)^\lambda A^\mu_{,\lambda} (x) + \frac{1}{2} (z-x)^\lambda (z-x)^\sigma A^{\mu}_{,\lambda\sigma} (x)\right]\nonumber$$
$$\left[A^\nu (y)  + (w - y)^\rho A^\nu_{,\rho} (y) + \frac{1}{2}(w-y)^\rho (w-y)^\kappa  A^{\nu}_{,\rho\kappa} (y)\right].\eqno(71)$$
The sorts of manipulation used in eqs. (18) and (19) then reduce eq. (69) to a form in which the integrations can be carried out and we find that
$$T_5 = \frac{i}{(4\pi it)^2} \int dx \left[\frac{2A^2}{(it)} + \left( -1 + 4(1-u) - 4(1-u)^2\right)A^\mu \partial_\mu \partial_\nu A^\nu\right.\nonumber$$
$$\left. + \left(2 (1-u) - 2(1-u)^2\right) A^\mu \partial^2 A_\nu\right].\eqno(72)$$
Last of all, for $T_6$ we find that insertion of complete sets of states leads to
$$T_6 = \int dx\, dy \delta (x-y) \int dp \,dq \int dz \frac{e^{ip\cdot (x-z)} e^{iq\cdot (z-y)}}{(2\pi)^8}\eqno(73)$$
$$\left[ e^{-i(1-u)p^2t}\left(g_{\mu\nu} - p_\mu p_\nu /p^2\right) + e^{-i(1-u)p^2t/\alpha} p_\nu p_\nu /p^2\right]
\left[ A^\mu (z) e^{-iuq^2t}A^\nu (y)\right].\nonumber$$
In eq. (73), $A^\mu (z)$ can now be expanded about $y$ up to second order in the derivatives of $A^\mu (y)$; following the steps outlined in computing $T_2 \cdots T_5$ above we arrive at
$$T_6 = \frac{i}{(4\pi it)^2} \int dx \left\lbrace \left[ \frac{3}{4} + \frac{1}{4(1-u + \frac{u}{\alpha})^2}\right] A^2 + (it) A_\mu \partial^2 A^\mu\right.\nonumber$$
$$\left. \left[ - \frac{5}{6} u^2 + \frac{3}{4} u - \frac{1}{6} \frac{u^2}{(1-u+\frac{u}{\alpha})^3} + \frac{1}{4} \frac{u}{(1-u+\frac{u}{\alpha})^2}\right]
+ (it) A^\mu \partial_\mu\partial_\nu A^\nu \left[ \frac{u^2}{3} - \frac{u^2}{3(1-u+\frac{u}{\alpha})^3}\right]\right\rbrace\ .\eqno(74)$$
Together, eqs. (72) and (74) reduce eq. (66) to
$$i S_A^{(1)} = \frac{i}{32\pi^2} \left\lbrace e^2 A^2 v^2 \left[ 3\lambda
\left(1 - \ln \frac{3\lambda v^2}{\kappa^2} \right) + (\lambda + \alpha e^2)\left( 1 - \ln \frac{(\lambda + \alpha e^2)v^2}{\kappa^2}\right)\right]\right.\nonumber$$
$$+ \int_0^1 du \left[ 
2A^2v^2 \left( (\lambda + \alpha e^2)(1-u) + 3\lambda u\right)
\left(1 - \ln \frac{((\lambda + \alpha e^2)(1-u) + 3\lambda u)v^2}{\kappa^2}  \right)\right.\nonumber$$
$$+ A^\mu \partial_\mu \partial_\nu A^\nu (-1 + 4u(1-u))
\left(- \ln \frac{((\lambda + \alpha e^2)(1-u) + 3\lambda u)v^2}{\kappa^2}  \right)\nonumber$$
$$\left. + A^\mu \partial^2 A_\mu (2u(1-u))
\left(- \ln \frac{((\lambda + \alpha e^2)(1-u) + 3\lambda u)v^2}{\kappa^2}  \right)\right]\eqno(75)$$
$$+ (2e^2 v)^2 \int_0^1 du \left[
A^2 \left(\frac{3}{4} + \frac{1}{4(1-u+ \frac{u}{\alpha})^2}\right)
\left(- \ln \frac{(e^2 (1-u) + 3\lambda u)v^2}{\kappa^2}  \right)
\right. \nonumber$$
$$+ \frac{1}{(e^2(1-u) + 3\lambda u)v^2} \left( A^\mu \partial^2 A_\mu
\left(- \frac{5}{6} u^2 + \frac{3}{4} u - \frac{1}{6} \frac{u^2}{(1-u+ \frac{u}{\alpha})^3}
\right.\right.\nonumber$$
$$\left. \left.\left.\left.
+\frac{u}{4} \frac{1}{(1-u+\frac{u}{\alpha})^2}\right)
+ A^\mu \partial_\mu \partial_\nu A^\nu \frac{u^2}{3} \left(1 - \frac{1}{(1-u + \frac{u}{\alpha})^3}\right)\right)\right]\right\rbrace.\nonumber$$
This is the contribution to the one loop effective action that is bilinear inthe vector field up to the second derivative of this field, in the gauge of eq. (31). Transversality is recovered in the limit $v \rightarrow 0$.

\section{Discussion}

We have demonstrated how operator regularization can be used to compute the expansion of the effective action in powers of derivates of the background field. This is illustrated in the context of a self interacting scalar model and in massless scalar electrodynamics.  In the latter calculation, the full dependence on the gauge parameter $\alpha$ of the two point function to second order in the derivatives of background fields has been determined. It is not clear how to reconcile the implied gauge dependence of the radiatively induced masses with the result of ref. [27-29].
Indeed the extension of the BRS identities given in refs. [29,30] and further extended in ref. [31] has been applied in refs. [32,33] to show that if eq. (26) holds, then $\frac{\partial V}{\partial\alpha}$ also vanishes in a class of gauges similar to those of eq. (31). In ref. (33), an expansion of the scalar mass $m^2$ given by eq. (27) was also shown to satisfy $\frac{\partial m^2}{\partial\alpha} = 0$. This expansion was taken to be $\left(\frac{\lambda v^2}{2}\right) + \sum^{(1)}(0) + 
\left(\frac{\lambda v^2}{2}\right) \sum^{(1)\prime}(0)$ where $\sum^{(1)}$ is the one-loop contribution to the scalar self-energy taken only to order $e^2\lambda$.
There is, however, an argument [34] that when one considers the exact effective potential in massless models, the effective potential is either background field independent or there is no spontaneous symmetry breaking.

There are alternate methods of making an expansion of the effective Lagrangian in powers of derivatives of the background field; these are the techniques of refs. [5-9] as well as these of refs. [35,36,37]. The approach of these last two references may well be more efficient than the one employed here.

\section{Acknowledgements}

The author would like to thank A. Buchel, V. Elias, V. Miransky and A. Rebhan for thoughtful comments. The hospitality of the Technical University of Vienna and the Perimeter Institute where much of this work was done is appreciated.  NSERC provided financial support. Roger Macleod made a useful suggestion.

\end{document}